\documentstyle[sprocl,epsf,12pt]{article}
\epsfverbosetrue
\def\PR #1 #2 #3 {Phys.~Rev.~{\bf #1}, #2 (#3)}
\def\PRL #1 #2 #3 {Phys.~Rev.~Lett.~{\bf #1}, #2 (#3)}
\def\PRD #1 #2 #3 {Phys.~Rev.~D~{\bf #1}, #2 (#3)}
\def\PLB #1 #2 #3 {Phys.~Lett.~{\bf B#1}, #2 (#3)}
\def\NPB #1 #2 #3 {Nucl.~Phys.~{\bf B#1}, #2 (#3)}
\def\RMP #1 #2 #3 {Rev.~Mod.~Phys.~{\bf #1}, #2 (#3)}
\def\ZPC #1 #2 #3 {Z.~Phys.~C~{\bf #1}, #2 (#3)}

\def\be{\begin{equation}}
\def\ee{\end{equation}}
\def\bea{\begin{eqnarray}}
\def\eea{\end{eqnarray}}

\begin{document}
\rightline{hep-ph/9608418}
\ \ \vskip3cm
\title{\bf TOP PHYSICS REVIEW AND OUTLOOK \footnote{Presented at the
XXXI Rencontres de Moriond on Electroweak Interactions and Unified Theories,
Les Arcs, Savoie, France, March 17-23, 1996.}
  }

\author{S.~WILLENBROCK}

\address {Department of Physics, University of Illinois,\\ 1110 West Green
Street \\  Urbana, IL, 61801}

\maketitle\vskip4in
\abstracts{I review the status of the top quark, and look forward to 
three topics relevant to future top-quark physics; spin correlation, 
single-top-quark production, and unification.}

\newpage
\setlength{\baselineskip}{1.50\baselineskip}
\section{Introduction}
\indent\indent In this talk I review the status of the top quark, 
and then look forward to the future of top-quark physics.  Top-quark 
physics is rich and varied, and I have chosen to present only a few 
specific topics which I find particularly interesting.  The topics to be 
covered are
\begin{itemize}

\item Top-quark yields

\item Top-quark mass and cross section

\item Future top-quark physics

\begin{itemize}

\item Spin correlation

\item Single-top-quark production

\item Top and unification

\end{itemize}

\end{itemize}
I end with a few concluding remarks.

\section{Top-quark yields}

\indent\indent Run I at the Fermilab Tevatron is now complete,
and each experiment has accumulated an integrated luminosity of about 
110 $pb^{-1}$.  The peak luminosity
achieved was about ${\cal L}=2 \times 10^{31}/cm^2/s$, impressive for
a machine that was designed for ${\cal L}=10^{30}/cm^2/s$.  The machine 
energy was 1.8 TeV.

Run II will begin in 1999, with a machine energy of 2 TeV.
The increase in energy is made possible by cooling the magnets to a lower
temperature, thereby allowing a higher field strength.  This increases
the top-quark production cross section by about $35\%$.

The most important change that will occur in Run II is a ten-fold
increase in luminosity, to ${\cal L}=2 \times 10^{32}/cm^2/s$.  This
will be achieved by two additions to the existing accelerator complex:
\begin{itemize}

\item Main Injector: The original Main Ring in the Tevatron collider
tunnel is a bottleneck to higher luminosity.  It will be replaced by the
Main Injector, a 120 GeV synchrotron housed in a separate tunnel, now
under construction.  The Main Injector will enable the production of many
more antiprotons, yielding a five-fold increase in luminosity.

\item Recycler: The Recycler ring is
an 8 GeV, low-field, permanent-magnet ring which will be installed in the
Main Injector tunnel.  The primary function of the Recycler is to allow
more efficient accumulation of antiprotons.  Its secondary role, from
which it takes its name, is to allow the reuse of antiprotons left over
from the previous store.  The Recycler will yield a two-fold increase in
luminosity.

\end {itemize}

The improvements in the accelerator, along with a variety of detector 
upgrades, 
will result in a dramatic increase in the potential for top-quark physics
in Run II.  In $t\bar t$ events, the final state with the most kinematic
information is
$W+4j$, where the $W$ is detected via its leptonic decay.  These events
are fully reconstructable. To reduce backgrounds, it is best to demand at
least one $b$ tag.  The number of such events is about 
500/$fb^{-1}$.\cite{TEV2000}$^)$ 
Depending on the length of Run II, the integrated
luminosity delivered to each detector will be between 1 and a few
$fb^{-1}$.  Thus there will be on the order of 1000 tagged,
fully reconstructed top-quark events in
Run II, to be compared with the approximately 25 $W+4j$ single-tagged top
events in Run I.

The CERN Large Hadron Collider (LHC) will provide another dramatic step 
forward in the potential for top-quark physics.  The increase in machine 
energy results in an increase in the top-quark production cross section 
of about a factor of 100.  The luminosity will range from 
$10^{33}-10^{34}/cm^2/s$.  Since $b$ tagging is so important to top-quark 
physics, it may be advantageous to study the top quark at the lower end 
of the luminosity range, where $b$ tagging is more effective.  The 
increase in cross section, together with the increase in luminosity, 
result in about a million top-quark pairs per year at the 
LHC.\cite{LHC}$^)$  The top-quark cross 
section and yield at the Tevatron and the LHC are summarized 
in Table 1.

\begin{table}[ht]
\begin{center}
\caption[fake]{Schedule, machine parameters, top-quark cross section, 
and top-quark yield, for Run II at the Tevatron and for the LHC.}
\bigskip
\begin{tabular}{ccc}
 & \underline{Tevatron Run II} & \underline{LHC} \\
\\
 & 1999 & 2004 \\
$\sqrt s$ & 2 TeV & 14 TeV \\
${\cal L}$ & $2 \times 10^{32}/cm^2/s$ & $10^{33}-10^{34}/cm^2/s$ \\
$\sigma_{t\bar t}$ & 6.5 $pb$ & 750 $pb$ \\
$W+4j$ ($b$ tag) & 500/$yr$ & $5 \times 10^{5}-10^{6}$/$yr$ \\
\end{tabular}
\end{center}
\end{table}

There are also possibilities for top-quark physics beyond the Tevatron 
and the LHC.  One proposal which is currently undergoing scrutiny is to 
increase the Tevatron luminosity yet further, to $10^{33}/cm^2/s$.  The 
``Tev33'', as it is called, might be useful for specialized top-quark 
studies, such as a measurement of the CKM matrix element $V_{tb}$ (more 
on this later).  

High-energy $e^+e^-$ and $\mu^+\mu^-$ colliders provide 
a complementary tool to hadron colliders for top-quark physics.  They 
have the unique capability of scanning the $t\bar t$ threshold region, 
but are also useful above the threshold region.  For lack of space, I 
will not be able to discuss top-quark physics at these colliders.

\section{Top-quark mass and cross section}

\indent\indent Preliminary values of the top-quark mass based on Run I 
data were presented at this conference by CDF and D0.  Averaging these 
two measurements, assuming they are uncorrelated, yields a world-average
top-quark mass of 
\begin{equation}
m_t=174.4 \pm 8.3 \;{\rm GeV}\;\;  {\rm (CDF/D0)}\;.
\end{equation}
It is anticipated that the error will be reduced to $\pm 4$ GeV in Run II.
The challenge to the Tev33 and the LHC is to go beyond this, perhaps to 
$\pm 1-2$ GeV.

The present measurement of the the $W$ mass is
\begin{equation}
M_W=80.330 \pm .150 \;{\rm GeV}\;\;  {\rm (UA2/CDF/D0)}\;.
\end{equation}
The error will be decreased in the near future by LEP II, and, somewhat 
later, by Run II at the Tevatron.  A world-average error of less than
$\pm 50$ MeV is anticipated.

\begin{figure}[htb]
\begin{center}
\epsfxsize= 0.8\textwidth
\leavevmode
\epsfbox{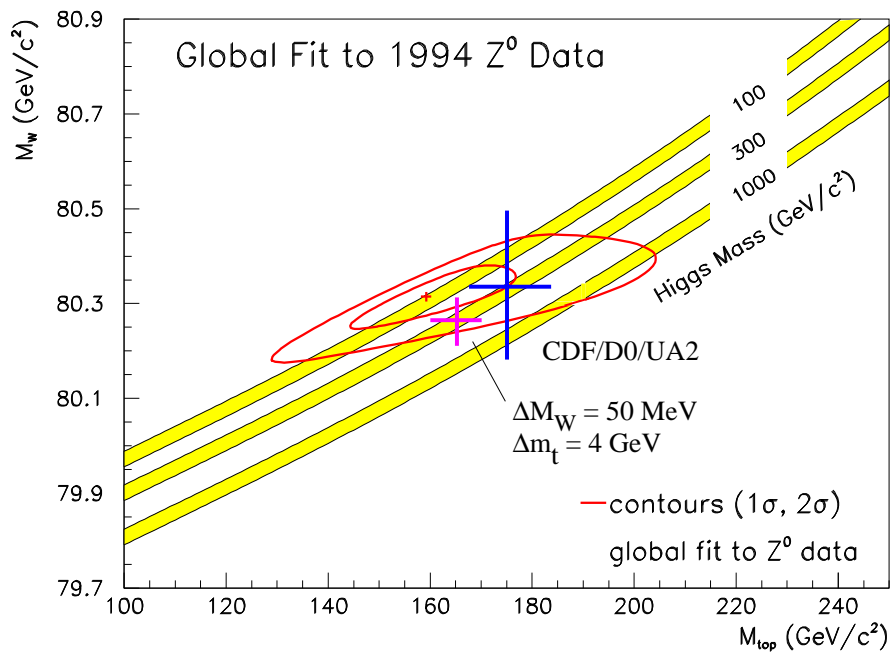}
\end{center}
\caption{$W$ mass vs.~top-quark mass, with bands of constant Higgs mass.
The contours are the one- and two-sigma regions from precision LEP and SLC
data.
The large cross is the direct measurement of $M_W$ and $m_t$.  The small cross,
placed arbitrarily on the figure, is the anticipated uncertainty in $M_W$ and
$m_t$ from LEP II and Run II at the Tevatron.}
\end{figure}

Figure~1 shows the well-known plot of $M_W$ vs.~$m_t$,
with bands of constant Higgs mass. The contours show the one- and
two-sigma fits to data from LEP and SLC, through 1994.\footnote{The 1995 
data is not yet available.}  The large cross indicates the
present direct measurements of $M_W$ and $m_t$.  The measurements are in 
good agreement with the precision electroweak data.
The small cross indicates the errors expected from LEP II and Run II at 
the Tevatron, placed arbitrarily on the plot. These measurements have the 
potential of indicating a preferred range for the Higgs mass, or of 
indicating physics beyond the standard Higgs model.

The production of top-quark pairs occurs via the 
strong processes $q\bar q, gg \to t\bar t$.  The top-quark cross section 
has been 
calculated at next-to-leading order in QCD.\cite{NDE,BKVS}$^)$  The most 
recent update of this calculation is shown in Fig.~2, as a 
function of the top-quark mass.\cite{CMNT}$^)$  The error band reflects the 
uncertainties in the uncalculated higher-order correction, in the 
parton distribution functions, and in $\alpha_s$.  
The average of the CDF and D0 cross sections reported at this conference, 
for a top-quark mass of 175 GeV, is also shown in the figure.  
The agreement between theory and experiment is quite satisfactory.

\begin{figure}[htb]
\begin{center}
\epsfxsize= 0.4\textwidth
\leavevmode
\epsfbox[25 145 530 660]{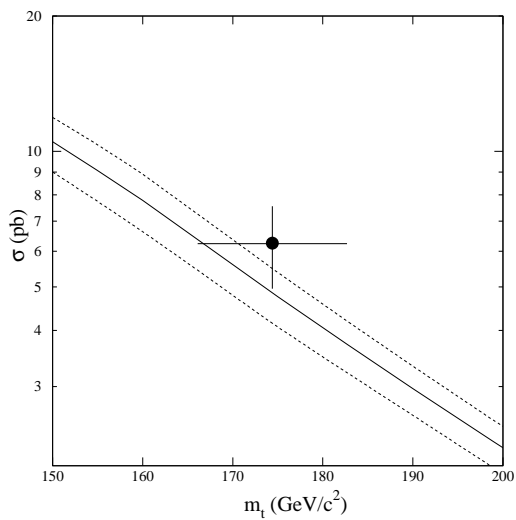}
\end{center}
\caption{Theoretical cross section, with error band, for $t\bar t$ 
production at the Tevatron ($\sqrt s=1.8$ TeV), versus the top-quark mass.
The CDF/D0 average mass and cross section are indicated.  Adapted from 
Ref.~5.}
\end{figure}

There are attempts to go beyond next-to-leading order in the theoretical 
calculation of the cross section.\cite{CMNT,LSV,BC}$^)$  Soft-gluon emission 
yields a correction 
proportional to $\alpha_s\ln^2 E/Q$, where $E$ is the gluon energy and $Q$ is 
the invariant mass of the $t\bar t$ pair.  This leads to a large correction to 
the cross section, proportional to $\alpha_s\ln^2 (s/4m_t^2-1)$, in the 
$t\bar t$ threshold region ($s\approx 4m_t^2$).   
Furthermore, these logarithmically-enhanced terms persist to all orders 
in perturbation theory, in the form $\alpha_s^n\ln^{2n} E/Q$.  
Fortunately, it is possible to sum these terms to all orders.  
However, there is no universally-accepted technique for performing 
this summation at this time.  Calculations lead to a correction, beyond 
next-to-leading order, of
$1\%$,\cite{CMNT}$^)$ $7\%$,\cite{LSV}$^)$ and $9\%$.\cite{BC}$^)$  The most 
striking aspect of this is that theorists are concerned with a correction 
of order $10\%$.  This is a great advance over the days, not so long ago, 
when QCD was considered successful if it agreed with data within $50\%$.

\section{Future top-quark physics}

\subsection{Spin correlation}

\indent\indent The top-quark lifetime is very short, 
$\Gamma_t^{-1}\approx (1.5 \;{\rm GeV})^{-1}$. 
This has the consequence that the top 
quark decays before the strong interaction has time to depolarize its 
spin.\cite{BDKKZ}$^)$
To understand this clearly, let's recall the situation for the $b$ quark, 
which is relatively long lived.  A $b$ quark hadronizes with a light 
antiquark into a $\overline{B}$ meson on a time scale 
$\Lambda_{QCD}^{-1}$.  Its spin is then depolarized on a time scale
$(\Lambda_{QCD}^2/m_b)^{-1}$, due to interactions of the light antiquark 
with the $b$-quark chromo-magnetic moment.  The $b$ quark decays, via the 
weak interaction, 
at a much later time.  For the top quark, these last two stages are
interchanged; the spin-depolarization time, 
$(\Lambda_{QCD}^2/m_t)^{-1}\approx (1.3 \;{\rm MeV})^{-1}$,
is three orders of magnitude longer than the top-quark lifetime.  Thus we 
expect the spin orientation of the top quark to be observable 
experimentally.

How does one go about testing this expectation?  Fortunately, the weak decay of 
the top quark is sensitive to its spin orientation; the angular 
distribution of the top-quark's decay products acts as a spin analyzer.  
Unfortunately, top quarks produced via the strong processes $q\bar 
q,gg\to t\bar t$ are unpolarized, because the strong interaction is 
parity conserving.  However, the spins of the $t$ and $\bar t$ are 
correlated.\cite{BOP,MP,SW1,B}$^)$  Thus if one observes this spin correlation 
experimentally, 
one has demonstrated that the top quark does indeed decay before the 
strong interaction has time to depolarize the top-quark's spin.

The differential cross section for $t\bar t$ production for different 
spin states is shown in Fig.~3, as a function of the $t\bar t$ invariant 
mass, at both the Tevatron and the LHC.  The subscripts $L$ and $R$ 
denote the helicity.  At the Tevatron, the $t$ and $\bar t$ 
are mostly produced with the opposite helicity, while at the LHC they 
tend to have the same helicity.  This difference is due to the fact that the 
dominant production process at the Tevatron is $q\bar q\to t\bar t$, 
while at the LHC it is $gg\to t\bar t$.  

\begin{figure}[htb]
\begin{center}
\epsfxsize= 0.5\textwidth
\leavevmode
\epsfbox[45 63 530 425]{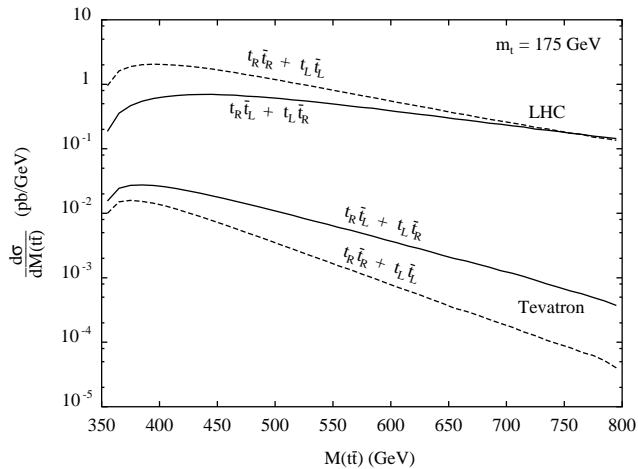}
\end{center}
\caption{Differential cross section for top-quark pair production, 
vs.~the $t\bar t$ invariant mass, for different helicity states, at the 
Tevatron and the LHC. From Ref.~11.}
\end{figure}

Since most of the total cross section comes from the threshold region, 
let's pause to reflect upon the origin of the spin correlation there.
Near threshold, the $t$ and $\bar t$ are produced in a state of zero 
orbital angular momentum.  For $q\bar q\to t\bar t$, which is 
mediated by an $s$-channel gluon, the $t\bar t$ is an ${}^3S_1$ state.
Two of these triplet states have the $t$ and $\bar t$ spins aligned; 
since the quarks recede from each other back-to-back, this results in 
them having the opposite helicity two-thirds of the time.  In the case of 
$gg\to t\bar t$, the $t\bar t$ is produced in a ${}^1S_0$ state, in which 
the spins are oppositely aligned, resulting in same-helicity $t\bar t$ 
pairs.

Although the spin correlation is large, its detection requires a significant 
amount of data.  It should be observable, at the $3\sigma$ level, with 1000 
$t\bar t$ events, as expected in Run II at the Tevatron.  The observation is 
aided by analyzing the spins along the beam axis (``beamline basis''), 
rather than along the 
direction of motion of the quarks (``helicity basis'').\cite{MP}$^)$
At the LHC, the spin 
correlation will be easily observable, and may be a useful tool to study 
the weak decay properties of the top quark.

\subsection{Single-top-quark production}

\indent\indent There are two processes which produce a single top quark, 
rather than a $t\bar t$ pair: the $W$-gluon-fusion process,\cite{DW,Y,EP}$^)$
depicted in Fig.~4(a), and $q\bar q \to t\bar b$,\cite{CP,SW2}$^)$
shown in Fig.~4(b). Both involve
the weak interaction, so they are suppressed relative to the strong
production of $t\bar t$; however, this suppression is partially compensated by
the presence of only one heavy particle in the final state.  Both processes
probe the charged-current weak interaction of the top quark.
The single-top-quark production cross
sections are proportional to the square of the Cabbibo-Kobayashi-Maskawa
matrix element $V_{tb}$, which cannot be measured in top-quark decays since
the top quark is so short-lived.

\begin{figure}[htb]
\begin{center}
\epsfxsize= 0.6\textwidth
\leavevmode
\epsfbox{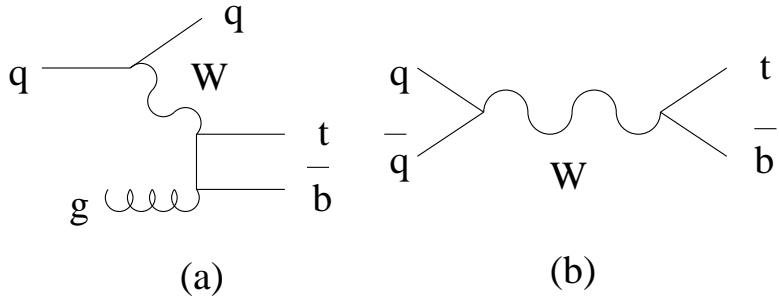}
\end{center}
\caption{Single-top-quark production at hadron colliders: (a) $W$-gluon fusion;
(b) quark-antiquark annihilation.}
\end{figure}

The cross sections for $t\bar t$, $W$-gluon fusion, and $q\bar q\to t\bar b$ 
are given in Table 2.  The process $q\bar q\to t\bar b$ is especially 
powerful as a measure of $V_{tb}$ due to its similarity to the 
Drell-Yan process, $q\bar q\to \ell\bar\nu$.  The Drell-Yan process can be 
used to help normalize the cross section, thereby reducing systematic 
uncertainties.  The parton distribution functions are relatively well 
known, in contrast with the gluon distribution function involved in the 
$W$-gluon fusion process.  At the LHC the process $q\bar q\to t\bar b$ is 
overwhelmed by backgrounds from $gg\to t\bar t$ and $W$-gluon fusion, 
since these 
processes are initiated by gluons.  Thus the Tevatron provides a better 
environment for the measurement of this process.

\begin{table}[ht]
\begin{center}
\caption[fake]{Approximate cross sections ($pb$) for top-quark production 
at the Tevatron ($\sqrt s=2$ TeV) and the LHC.}
\bigskip
\begin{tabular}{ccc}
 & \underline{Tevatron} & \underline{LHC} \\
\\
$t\bar t$ & 6.5 & 750 \\
$Wg\to t\bar b$ & 2 & 200 \\
$q\bar q\to t\bar b$ & 0.88 & 10 \\
\end{tabular}
\end{center}
\end{table}

The cross section for $q\bar q\to t\bar b$ has been 
calculated at next-to-leading order in QCD.\cite{SmithW}$^)$
The dominant sources of 
uncertainty are the parton distribution functions and the top-quark mass.
A measurement of $V_{tb}$ to an accuracy of $10\%$ should be possible 
in Run II at the Tevatron, and perhaps to $4\%$ with Tev33, assuming 
$V_{tb}$ is close to unity.  

\subsection{Top and unification}

\indent\indent The top quark is much more massive than the other known 
fermions, and 
this may provide a clue towards understanding nature at a deeper level. 
Nature has encouraged us to extrapolate the gauge couplings to high 
energy, where they are successfully unified (with the additional 
assumption of weak-scale supersymmetry) into a single SU(5) gauge 
coupling, $g_U$, at the GUT scale, $\sim 10^{16}$ GeV.  The value of $g_U$ is 
approximately $1/\sqrt 2$, a number of order unity, as one would expect in a truly
fundamental description of nature.

In supersymmetric models, fermion masses arise from their Yukawa 
coupling to one of the two Higgs fields, similar to the standard Higgs 
model (which has just one Higgs field).  The Yukawa coupling of the top quark 
is of order unity at the weak scale.  
When extrapolated up to the GUT scale, it remains a number of order 
unity, which is encouraging.  However, if the top quark were a bit 
heavier, its Yukawa coupling would blow up before reaching the GUT scale.  
Thus there is an upper bound of about 200 GeV on the top-quark mass in 
supersymmetric GUT models.\cite{BDM}$^)$  The fact that nature has chosen not 
to provide us with a quark heavier than this bound 
further encourages us to pursue supersymmetric grand unification.

The other fermions have very small Yukawa couplings, which is difficult 
to understand from a fundamental perspective.  To a good approximation, 
they are zero in comparison with the top-quark's Yukawa coupling.  An 
appealing explanation for this arises naturally in string theory.  String 
theories are replete with discrete symmetries, and these symmetries 
make it difficult to have Yukawa couplings.  The best that 
one can usually achieve, in the context of three generations of quarks 
and leptons, is one and only one nonvanishing Yukawa coupling, of order 
unity.\cite{F,CHL}$^)$  The hope is that the other Yukawa couplings, which are 
zero at leading order, arise from small perturbations to this scenario.  

\section{Conclusions}

\indent\indent We are presently in the dawn of the top-physics era.  The 
future promises a wealth of top physics at the Tevatron, LHC, 
and perhaps high-energy $e^+e^-$ and $\mu^+\mu^-$ colliders.  The large 
top-quark mass allows for accurate perturbative calculations of electroweak 
and strong top-quark processes.  The experimental challenge 
is to match and surpass the accuracy of these calculations, in order to test 
the properties of the top quark with the greatest possible sensitivity.  
Since the top quark is by far the heaviest fermion, it would be a mistake 
to assume that its properties are simply those given by the standard model.
Perhaps the top quark is exotic in some way, and will give us 
our first glimpse of physics beyond the standard model.

\section*{Acknowledgements}

\indent\indent I am grateful for conversations with and assistance from
G.~Anderson, T.~Liss, J.~Lykken, R.~Roser, and T.~Stelzer.
This work was supported in part by Department of Energy grant
DE-FG02-91ER40677.

 \clearpage

\end{document}